\documentclass{article}
\usepackage{graphicx}
\graphicspath{{converted_graphics/}}
\begin{document}

\begin{center}
{\Large Measurements of Ice Crystal Growth Rates in Air at -5C and -10C}%
\vskip9pt

{\Large K. G. Libbrecht and H. M. Arnold}\vskip4pt

{\large Department of Physics, California Institute of Technology}\vskip-1pt

{\large Pasadena, California 91125}\vskip-1pt

{\large address correspondence to: kgl@caltech.edu}\vskip-1pt

\vskip18pt

\hrule\vskip1pt \hrule\vskip14pt
\end{center}

\noindent \textbf{Abstract. We present experiments investigating the growth
of ice crystals from water vapor in air using a free-fall convection
chamber. We measured growth rates at temperatures of -5 C and -10 C as a
function of supersaturation at an air pressure near one bar. We compared our
data with numerical models of diffusion-limited growth based on cellular
automata to extract surface growth parameters at different temperatures and
supersaturations. From these investigations we hope to better understand the
surface molecular dynamics that determine crystal growth rates and
morphologies.}

\vskip4pt \noindent \textit{[The figures in this paper have been reduced in
size to facilitate rapid downloading. The paper is available with higher
quality figures at http://www.its.caltech.edu/\symbol{126}%
atomic/publist/kglpub.htm, or by contacting the author.]}

\section{Introduction}

The formation of complex structures during solidification often results from
a subtle interplay of nonequilibrium, nonlinear processes, for which
seemingly small changes in molecular dynamics at the nanoscale can produce
large morphological changes at all scales. One popular example of this
phenomenon is the formation of snow crystals, which are ice crystals that
grow from water vapor in a background gas. Although this is a relatively
simple physical system, snow crystals display a remarkable variety of
columnar and plate-like forms, and much of the phenomenology of their growth
remains poorly understood \cite{libbrechtreview}.

Recent experimental and theoretical work suggests that surface impurities
may play an essential role in determining snow crystal growth rates and
morphologies under normal atmospheric conditions \cite{impurities}. To
investigate this further we need precision measurements of snow crystal
growth dynamics over a range of conditions, especially as a function of the
type and concentration of active impurities within an inert background gas.
We have constructed a free-fall convection chamber for making such
measurements \cite{chamber}, and we recently described ice growth data
obtained over a range of temperatures using this chamber \cite{2008data}.
The present paper describes additional experiments using the same apparatus,
in which we measured growth rates as a function of supersaturation at -5 C
and -10 C.

\section{Observations and Modeling}

Our experiments were performed in a convection-mixed ice crystal growth
chamber containing ordinary air at atmospheric pressure \cite{chamber}. We
used a heated reservoir filled with deionized water inside the chamber to
produce a known water vapor supersaturation via evaporation and convective
mixing. A number of crystals were nucleated and allowed to grow for several
minutes while in free-fall inside the chamber. At various times we briefly
opened a shutter that allowed crystals to fall onto a glass substrate at the
bottom of the chamber, where we measured their size and thickness using a
combination of optical imaging and broad-band interferometry \cite{chamber}.
From observations of a large number of crystals we obtained the average
crystal dimensions as a function of growth time under conditions of known
temperature and supersaturation, as well as some sense of the distribution
of these quantities. Because of outgassing from the chamber walls and other
sources, we expect that the background air in our chamber included a number
of unknown impurities at the part-per-million level. We now suspect that
these impurities may have a substantial effect on the ice growth dynamics 
\cite{impurities}, and we will be investigating this hypothesis further in
future experiments.

\subsection{Measurements at -5C}

\begin{figure}[ht] 
  \centering
  \includegraphics[bb=0 0 661 540,width=3.3in,height=2.7in,keepaspectratio]{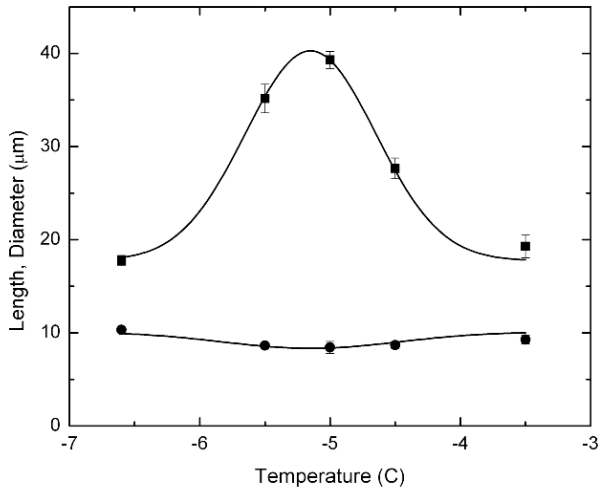}
  \caption{Measurements of the average
lengths (top points) and diameters (lower points) of columnar crystals after
120 seconds of growth at a supersaturation of 2.3 percent. The upper curve
shows a fit Gaussian centered at -5.15 C.}
  \label{needlepeak}
\end{figure}

Our first data were taken near the needle peak in the snow crystal
morphology diagram \cite{libbrechtreview}. To accurately locate the position
of this peak as a function of temperature, we examined crystals grown for
120-180 seconds at different temperatures in air with a supersaturation of $%
\sigma _{\infty }=2.3$ percent. The crystal dimensions were approximately
proportional to growth time in these measurements, so we fit the data to
determine average crystal dimensions after 120 seconds of growth. Results
are shown in Figure \ref{needlepeak}, and a fit to the column length data
yields a peak at $T\approx -5.15$ C.

\begin{figure}[p] 
  \centering
  \includegraphics[bb=0 0 2024 3509,width=4.0in,keepaspectratio]{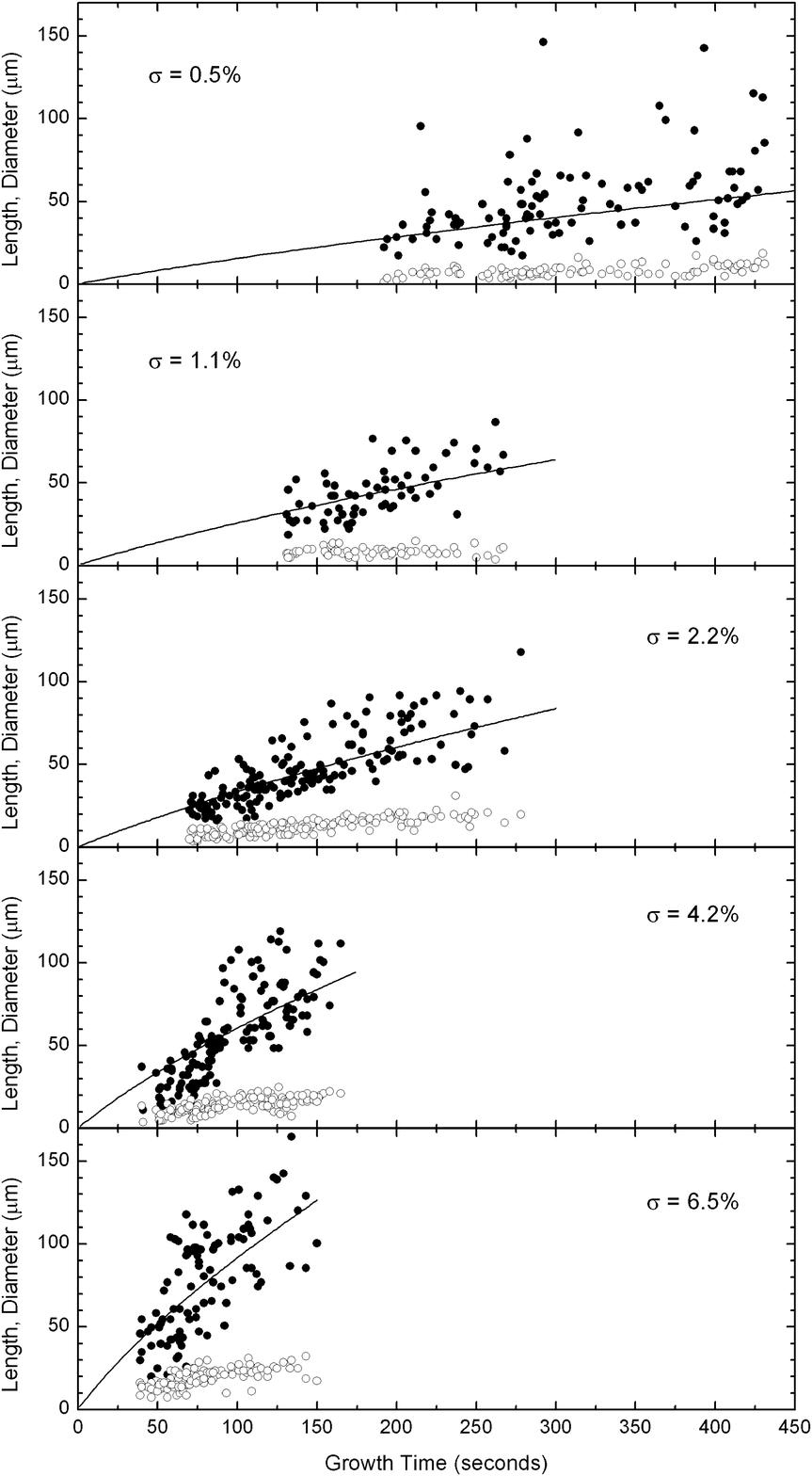}
  \caption{Growth of columnar crystals at $%
T=-5.0\pm 0.15$ C as a function of time for the five different values of $%
\protect\sigma _{\infty }$ indicated, showing column lengths (filled points)
and diameters (open points). The lines are numerical models of
diffusion-limited growth described in the text.}
  \label{length5c}
\end{figure}

\begin{figure}[p] 
  \centering
  \includegraphics[bb=0 0 1543 2639,width=4.0in,keepaspectratio]{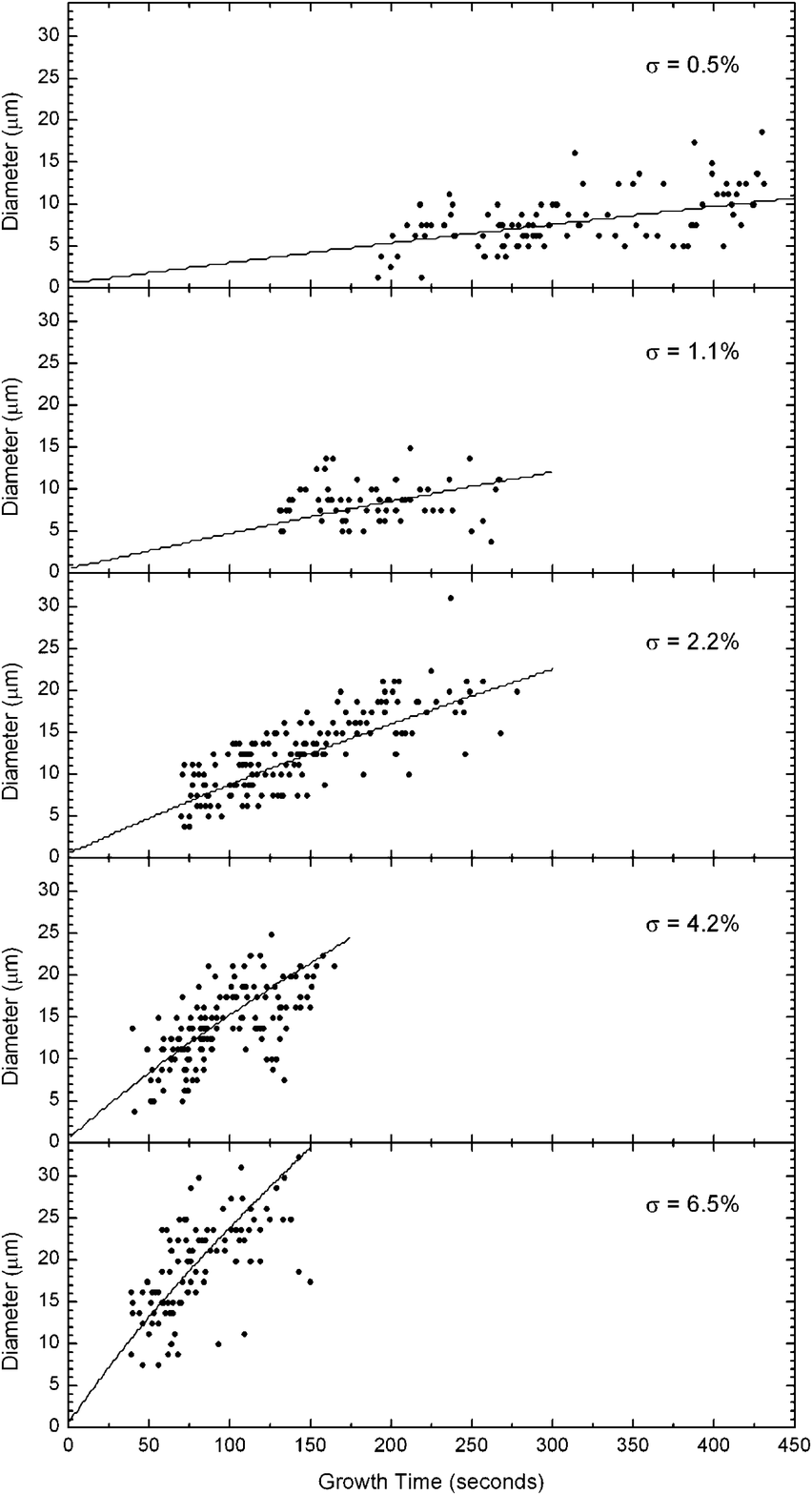}
  \caption{Same data as in Figure \protect\ref%
{length5c}, but here showing just diameters with an expanded scale. Again
the curves show numerical models described in the text.}
  \label{diam5c}
\end{figure}

We also observed crystals growing at $T=-5.0\pm 0.15$ C at different
supersaturations. Once the growth chamber had stabilized, these data were
taken by: 1) nucleating crystals near the top of the chamber as described in 
\cite{chamber}; 2) waiting some length of time with the shutter closed (so
crystals could not fall on the substrate); 3) opening the shutter; and 4)
visually scanning the substrate and recording crystals for about 30 seconds.
A typical cycle yielded about 5-10 crystal measurements. Using different
wait times before opening the shutter, we were able to obtain crystal sizes
as a function of growth time (equal to time after nucleation). Results are
shown in Figures \ref{length5c} and \ref{diam5c}.

\begin{figure}[t] 
  \centering
  \includegraphics[bb=0 0 1324 1108,width=3.5in,keepaspectratio]{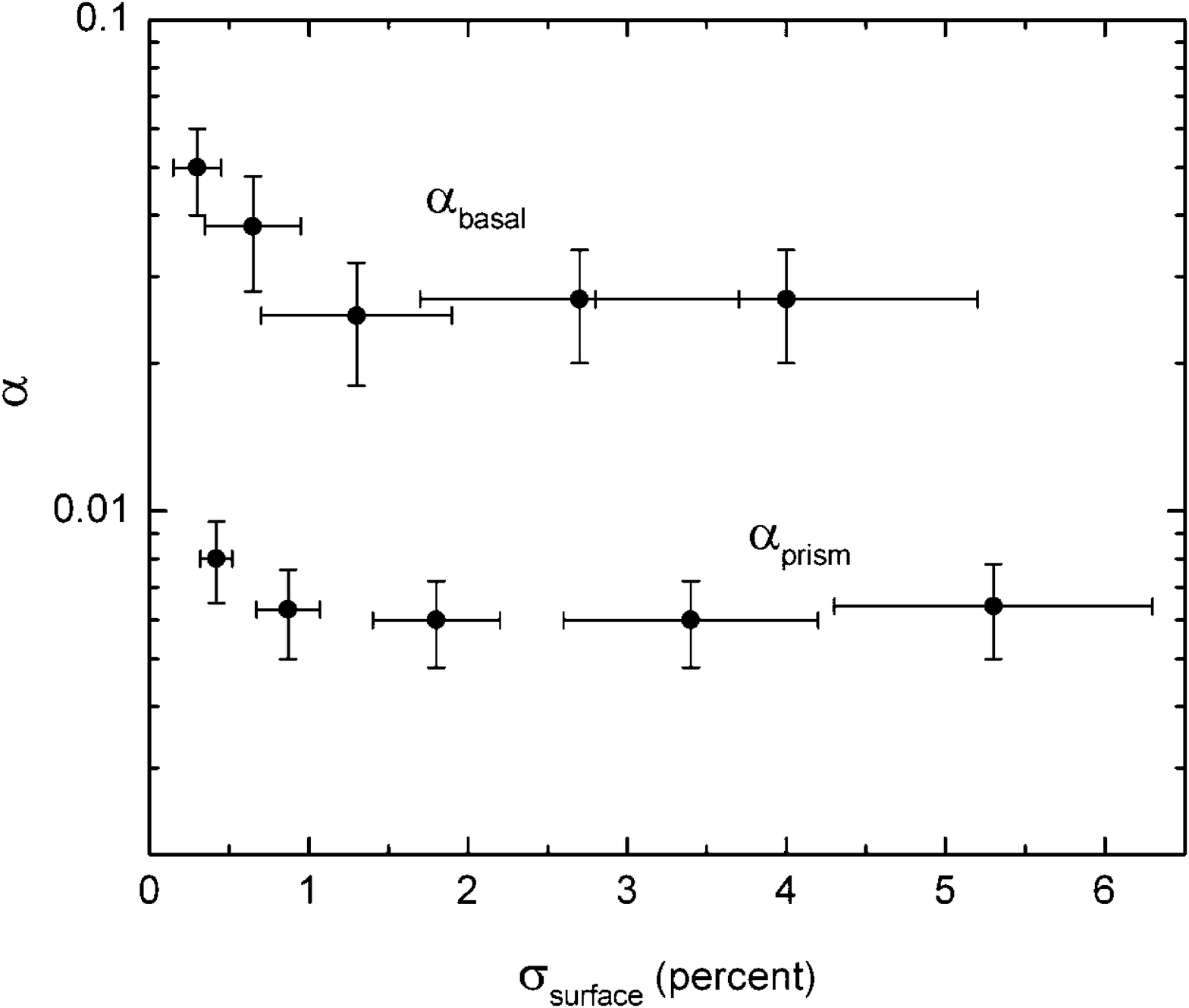}
  \caption{Attachment coefficients for the prism and basal facets as a function
of surface supersaturation for columnar snow crystals grown at -5 C.}
  \label{alphas5}
\end{figure}

We compared our measurements to cylindrically symmetric numerical models of
diffusion-limited growth using a cellular automata method \cite%
{libbrechtmodel, 2008data}. Input to the models included the attachment
coefficients $\alpha _{prism}$ and $\alpha _{basal}$ for the two crystal
facets, and these parameters were adjusted to fit the data. The initial
crystal size in all cases was 1 $\mu $m. From the best-fit models we
extracted $\sigma _{surface},$ the supersaturation at the crystal surface,
as a function of time and location on the surface. For each model we
estimated an average value of $\sigma _{surface}$ and the variation in that
value, and we interpreted the latter as an estimate of uncertainty in the
inferred $\sigma _{surface}.$ Figure \ref{alphas5} shows results from this
modeling of the data.

\subsection{Measurements at -10C}

\begin{figure}[p] 
  \centering
  \includegraphics[bb=0 0 1845 2206,width=5in,keepaspectratio]{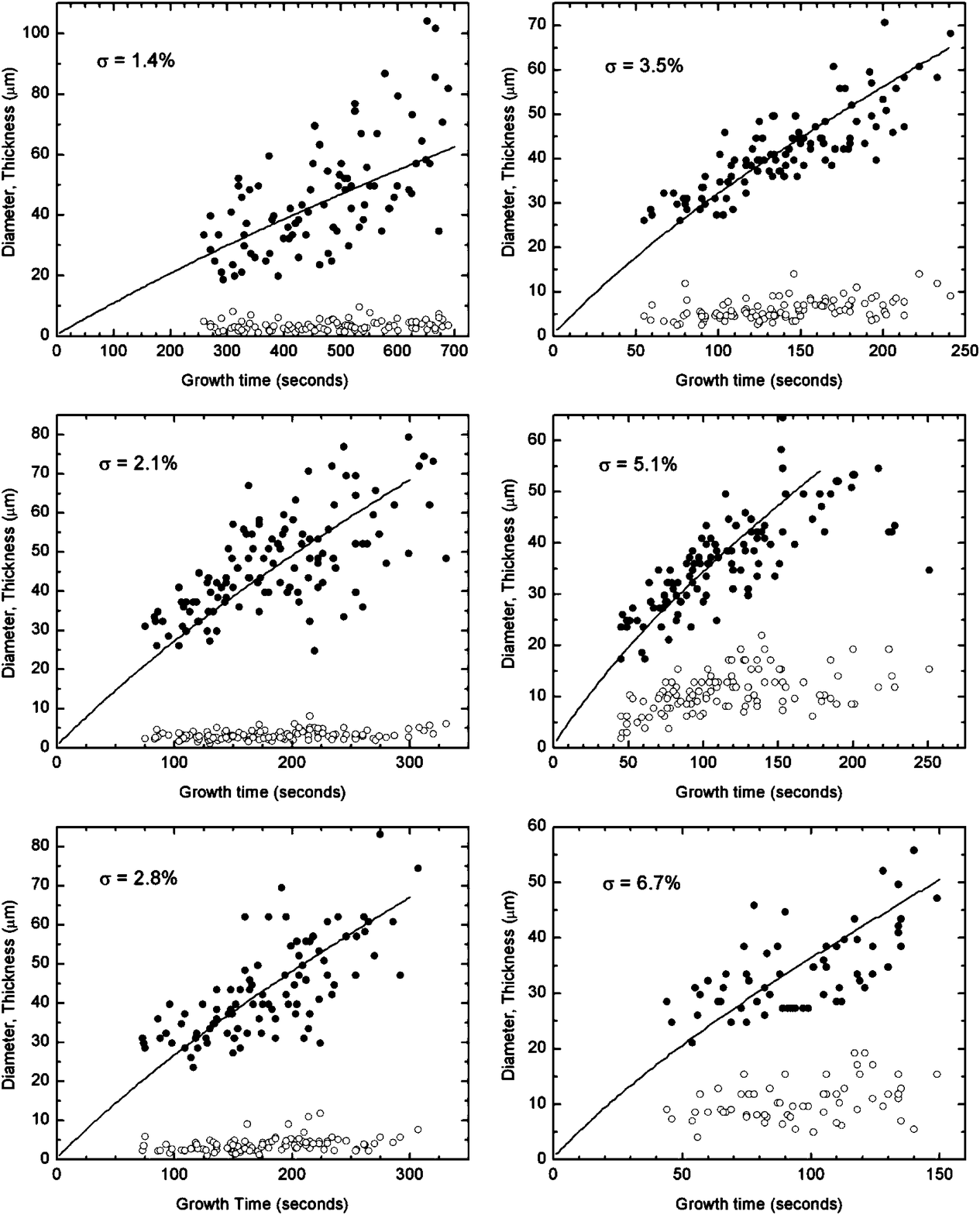}
  \caption{Growth of simple plate-like
crystals at $T=-10.0\pm 0.2$ C as a function of time for the six different
values of $\protect\sigma _{\infty }$ indicated, showing plate diameters
(filled points) and thicknesses (open points). The lines are numerical
models of diffusion-limited growth described in the text.}
  \label{diams10}
\end{figure}

\begin{figure}[p] 
  \centering
  \includegraphics[bb=0 0 2023 2478,width=5in,height=5.9in,keepaspectratio]{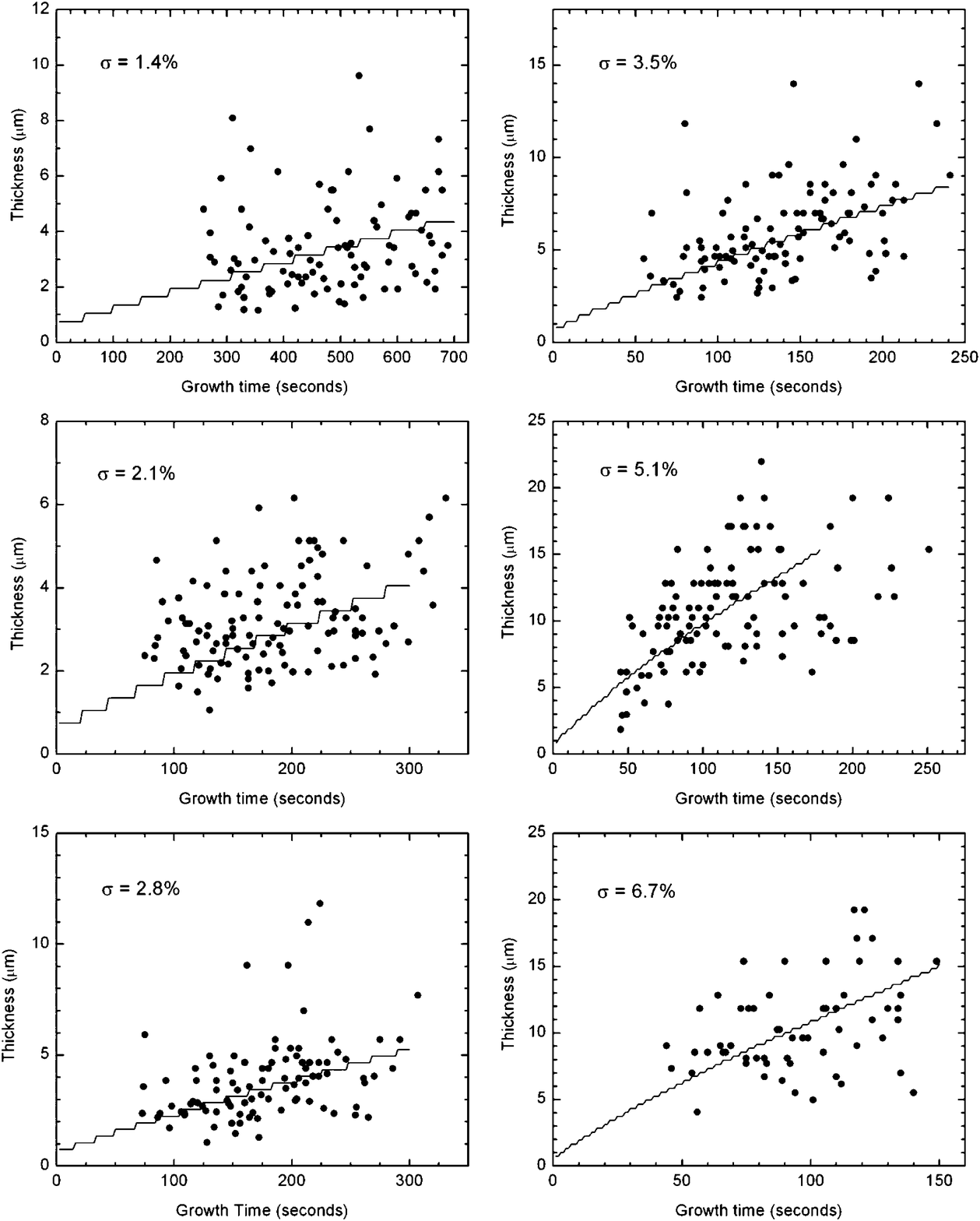}
  \caption{Same data as in Figure \protect\ref{diams10}, but here showing
just thicknesses of the plate-like crystals, with an expanded scale. Again
the curves show numerical models described in the text.}
  \label{thick10}
\end{figure}

\begin{figure}[th] 
  \centering
  \includegraphics[bb=0 0 905 739,width=3.1in,height=3.35in,keepaspectratio]{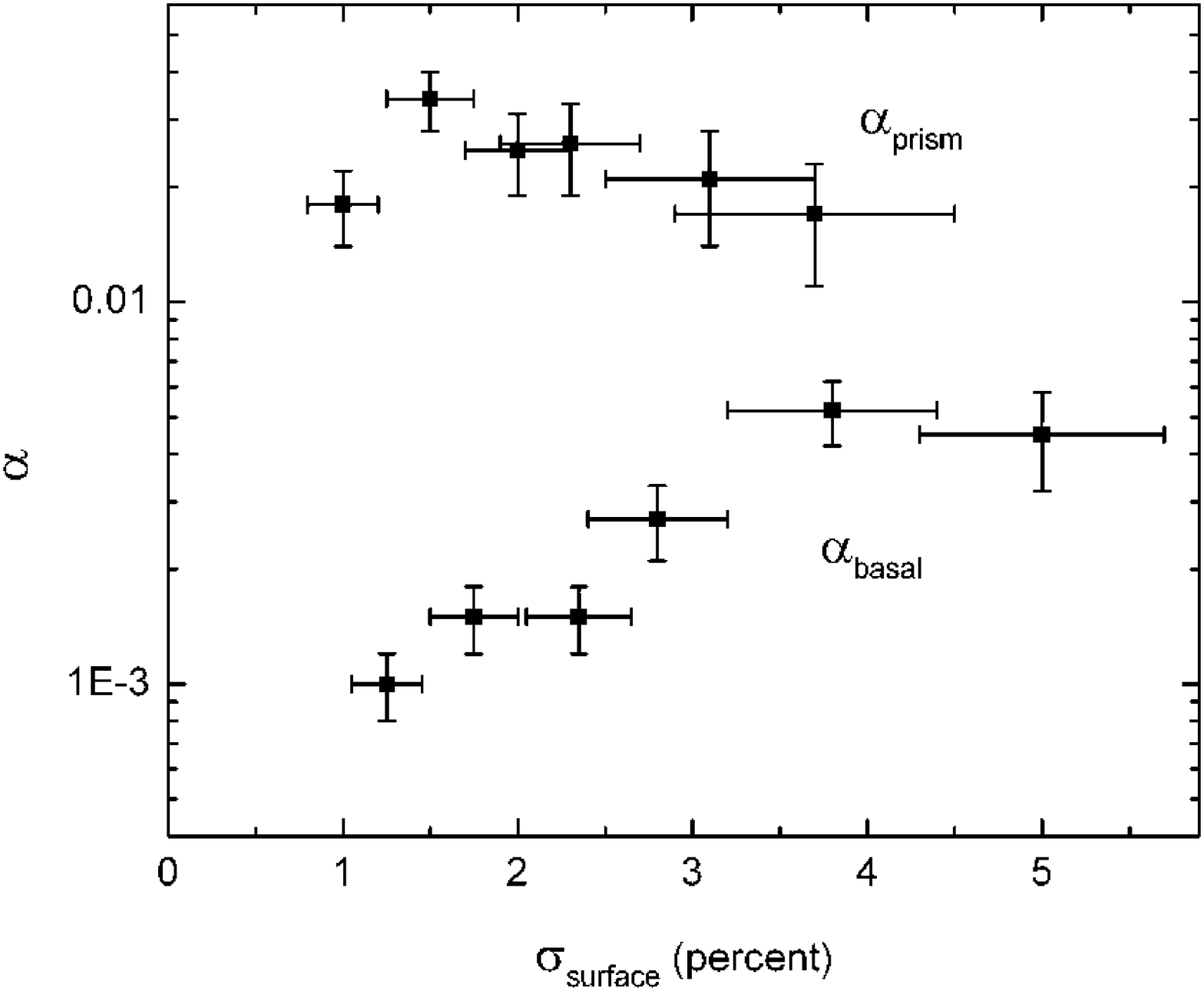}
  \caption{Attachment coefficients for the prism and basal facets
as a function of surface supersaturation for plate-like snow crystals grown
at -10 C.}
  \label{alphas10}
\end{figure}

We obtained a second series of measurements at $T=-10.0\pm 0.2$ C, following
the same procedures described above for the -5 C measurements, and the raw
data are shown in Figures \ref{diams10} and \ref{thick10}. Here we define
the effective diameter of a plate-like crystal as $D=(4A/\pi )^{1/2}$, where 
$A$ is the projected basal area of the plate. Most crystals in these data
were simple hexagonal plates. Numerical modeling again yielded the
attachment coefficients $\alpha _{basal}$ and $\alpha _{prism}$ as a
function of $\sigma _{surface}$ at $T=-10$ C, shown in Figure \ref{alphas10}.

\section{Possible Systematic Errors}

There are a host of experimental and modeling difficulties associated with
measuring ice crystal growth rates and determining surface attachment
coefficients, and many earlier experiments have been affected by a variety
of systematic errors \cite{errors}. We have made substantial efforts to
minimize these problems in the present experiments, but it is nevertheless
useful to examine a number of possible remaining systematic errors in detail.

\subsection{Defining the Environment}

We determined the temperature and supersaturation in our growth chamber
following the procedures described in \cite{chamber, 2008data}. Convective
mixing created a surprisingly uniform temperature inside the chamber, even
in the presence of the heated water reservoir \cite{chamber}. By direct
measurement we found that the air temperature in the chamber was uniform
over most of its volume to $\pm 0.15$ C, and we measured the temperature
with an accuracy of $\pm 0.1$ C. The supersaturation $\sigma _{\infty }$
depended on the temperature of the water reservoir, and we described
measurements for calibrating $\sigma _{\infty }(T_{water})$ in \cite{chamber}%
. We estimate that $\sigma _{\infty }$ is known in the present experiments
to approximately $\pm 20$ percent. In addition to random errors, we expect
there are likely systematic trends in $\sigma _{\infty ,calculated}-\sigma
_{\infty ,actual}$ as a function of $\sigma _{\infty }$ at the $\pm 20$
percent level.

\subsection{Competition Effects}

\begin{figure}[ht] 
  \centering
  \includegraphics[bb=0 0 752 612,width=3.2in,keepaspectratio]{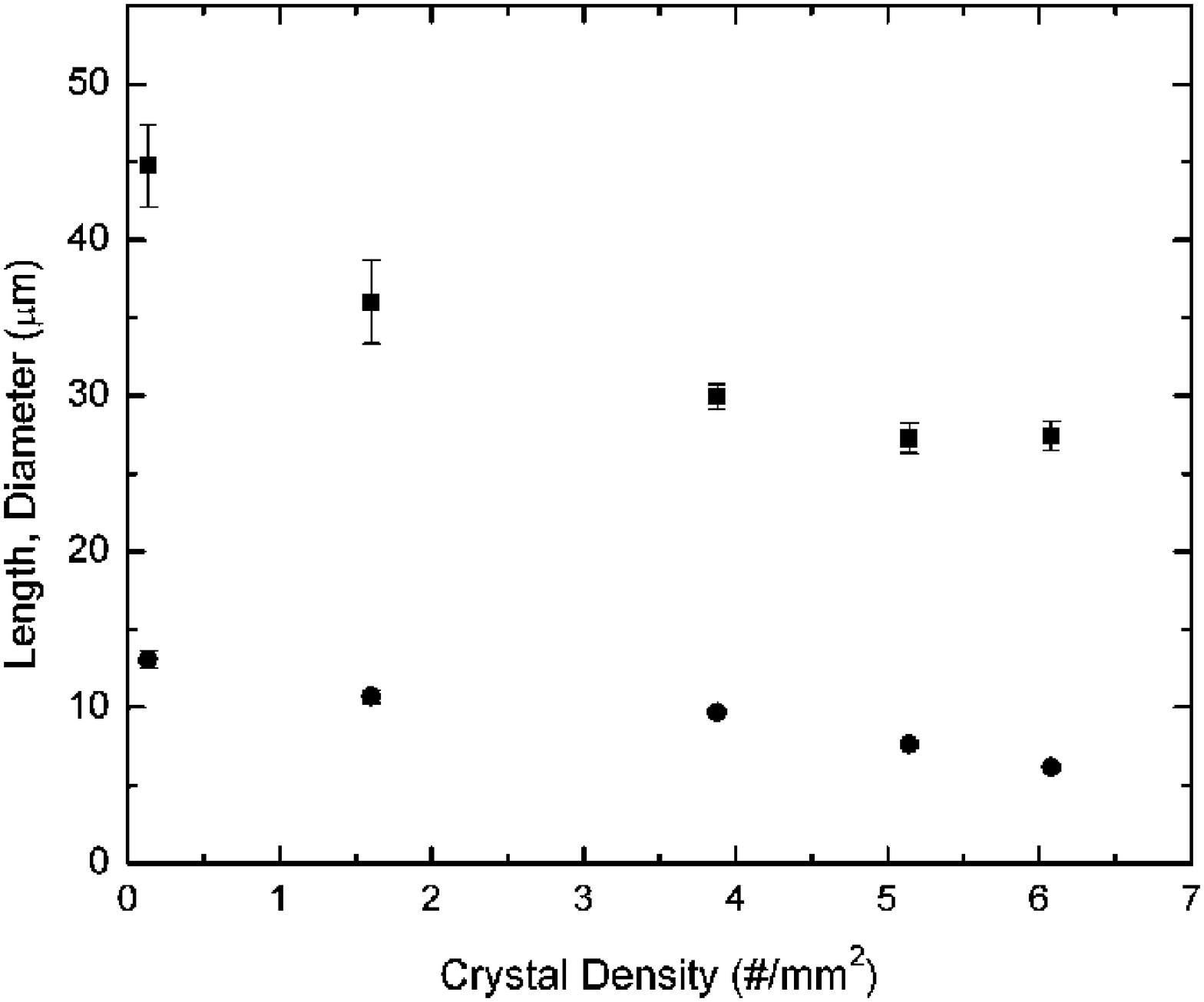}
  \caption{Average lengths (top points) and
diameters (lower points) of columnar crystals growth at -5 C, as a function
of the number density of crystals that fell onto the substrate. Points are
from nucleation pressures of 12, 14, 16, 18, and 20 psi (from left to
right). This plot shows that the supersaturation is not greatly affected by
the presence of growing crystals in our measurements. We typically operated
with low nucleation pressures to minimize this systematic error.}
  \label{nucleation}
\end{figure}

Our calibration of $\sigma _{\infty }$ was done with no crystals freely
falling in the chamber \cite{chamber}, while our measurements were made with
an undetermined number of growing crystals in the chamber. One possible
systematic error arises if the number of crystals in the chamber is so large
that they collectively remove water vapor from the air faster than it can be
replenished, thus effectively reducing $\sigma _{\infty }.$ We tested this
by making measurements using different nucleation pressures (see \cite%
{chamber}) to produce different numbers of crystals in the chamber. Figure %
\ref{nucleation} shows average crystal sizes as a function of the density of
crystals found on the substrate after a fixed time. These data were taken at
a temperature of $T=-5$ C and a background supersaturation of $\sigma
_{\infty }=4.5$ percent.

Extrapolating the points in Figure \ref{nucleation} to zero density would
give the limit in which the growing crystals have no effect on the ambient
supersaturation in the chamber. We typically operate with low nucleation
pressure, essentially at the left-most point in the Figure. The data thus
suggest that our inferred supersaturation is not reduced more than perhaps
20 percent by the presence of growing crystals.

\subsection{Sampling Errors}

Another potential systematic error occurs because crystals are not observed
immediately after they land on the substrate, because it takes time to find
them as the substrate is scanned. Thus each measurement of a crystal size
gives $S(t_{growth}+t_{wait}),$ where $t_{growth}$ is the time between
nucleation and when the crystal landed on the substrate, and $t_{wait}$ is
the time between when the crystal landed and when it was observed. In the
present measurements we opened the shutter and scanned the substrate for
about 30 seconds, thus ensuring that $t_{wait}<30$ seconds, and we estimate
that typical values were $t_{wait}\approx 10$ seconds.

In our previous measurements we reported evidence for an initial rapid
crystal growth followed by slower growth at $T=-10$ C \cite{2008data}. More
careful measurements in the present experiments did not confirm this
behavior, and we now believe that sampling errors were distorting our
earlier data. In short, we sampled the substrate too long while collecting
the -10 C data in \cite{2008data}, and the long $t_{wait}$ produced a signal
that mimicked an initial rapid growth period.

\subsection{Sedimentation Effects}

Our crystals were grown while in free fall inside our chamber, and observed
once they landed on the substrate. Thus we only sample crystals that make it
to the substrate, which is not a completely unbiased sample. At early times,
for example, we expect that the larger and heavier crystals will be
oversampled at the substrate. Likewise, at later times most of the crystals
will have already fallen, so we will be sampling crystals that fall more
slowly than average. We do not believe this was a large effect, but it may
have distorted our data somewhat. In our modeling, we tended to give less
weight to crystals that fell especially early or especially late in our
measurements.

\subsection{Substrate Growth Effects}

Once a crystal lands on our substrate, it will grow or evaporate with time,
depending on the temperature of the substrate relative to the surrounding
air. This behavior was easily seen in our experiments, so we took
considerable care to minimize any systematic errors that resulted. First, we
carefully controlled the substrate temperature and operated near the
equilibrium point, where the crystals did not grow or evaporate. And second,
we kept $t_{wait}$ small to minimize growth/evaporation effects.
Occasionally we made observations as a function of $t_{wait}$ to estimate
the size of any residual systematic errors. We believe that substrate growth
effects did not alter the measurements more than roughly $\pm 20$ percent.

\subsection{Modeling Errors}

Once we have produced trustworthy crystal size measurements as a function of
growth time, there are additional errors that may come in when modeling the
growth data to yield attachment coefficients. These are much reduced if the
observed crystals are small, and they are also reduced when the attachment
coefficients are small \cite{libbrechtreview}. Then the growth is limited
mainly by attachment kinetics, so there is little difference between $\sigma
_{surface}$ and $\sigma _{\infty }$, and the attachment coefficients are
easily extracted from the growth data for simple crystals \cite%
{libbrechtreview}. In our data the differences between $\sigma _{surface}$
and $\sigma _{\infty }$ can be seen by comparing the numbers in Figures \ref%
{diam5c} and \ref{alphas5} for the -5 C data, and likewise from Figures \ref%
{thick10} and \ref{alphas10} for the -10 C data. Because the differences are
fairly small, we expect that errors introduced by imperfections in our
modeling of diffusion-limited growth are likely not substantial.

There is one exception, however, in our inference of $\alpha _{basal}$ in
the -5 C data. We observed a substantial degree of hollowing in the columnar
crystals at -5 C, especially at the higher supersaturations. Our model
crystals, however, showed little hollowing at all supersaturations. The
difference probably means that our inferred $\alpha _{basal}$ at -5 C are
too high, especially at the higher supersaturations, but even this statement
is uncertain. We plan to address the hollowing issue in a future
investigation.

It is instructive to compare the present results with similar data from \cite%
{yu}, since both experiments included measurements of freely falling
crystals in air near $-5$ C. The raw data in \cite{yu} compare reasonably
well with the data presented here, and yet the conclusions of the two
studies are quite different: Figure \ref{alphas5} shows little variation in $%
\alpha _{prism}$ and $\alpha _{basal}$ with $\sigma _{surface},$ whereas
Figure 4 in \cite{yu} shows large reductions in both $\alpha _{prism}$ and $%
\alpha _{basal}$ for $\sigma _{surface}<1$ percent. We investigated this
discrepancy and found that most of the difference comes from modeling.
First, the Green's function technique used in \cite{yu} is not nearly as
stable as the cellular-automata method used here, and we believe that the
difference introduced errors in the inferred attachment coefficients and
surface supersaturations in \cite{yu}. Second, the power-law fits assumed in 
\cite{yu} were inappropriate. This functional form fits the data reasonably
well only because we have so little data at early times. The cellular
automata models shown in Figures \ref{length5c} and \ref{diam5c}, however,
are not well described by the power-law behavior assumed in \cite{yu}. We
were also more careful to record an unbiased sample of growing crystals in
the present experiments, but this did not seem to make a large difference
compared to the data in \cite{yu}.

\section{Discussion}

\begin{figure}[ht] 
  \centering
  \includegraphics[bb=0 0 1855 1477,width=4.58in,height=3.65in,keepaspectratio]{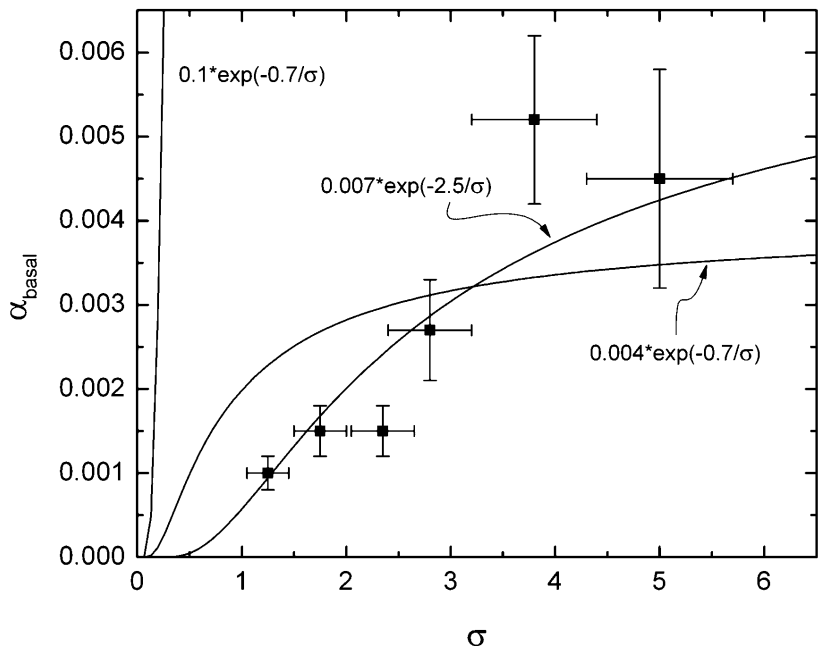}
  \caption{Comparison of the present data
for $\protect\alpha _{basal}(\protect\sigma _{surface})$ at -10 C with
different models, as described in the text.}
  \label{summary10}
\end{figure}

Figure \ref{summary10} shows a comparison of the present data at -10 C with
crystal growth data taken at the same temperature but in a low-pressure
environment \cite{precision}. The highest curve in Figure \ref{summary10}
shows a fit to measurements of $\alpha _{basal}$ at 0.005 bar, showing that $%
\alpha _{basal}$ is much higher at low pressure. We have speculated
previously that this difference arises from surface impurities impeding ice
growth in the 1-bar data \cite{impurities}. The other two curves show that
using $\sigma _{0}=0.7$ percent (from the low-pressure data \cite{precision}%
) yields a poor fit to the 1-bar data, and that the latter suggest a value
closer to $\sigma _{0}=2.5$ percent. The assumption of 2D-nucleation-limited
growth may not be a good one in the presence of surface impurities, however,
and some other growth model may be needed.

At present we have essentially no theory to explain the phenomenology of
snow crystal growth as a function of temperature and supersaturation \cite%
{libbrechtreview}. Basic atomistic crystal growth theory is simply not
sufficient to explain the observations, even at a qualitative level. The
data presented here and in \cite{2008data} are but a first step toward
collecting precise measurements of $\alpha _{prism}$ and $\alpha _{basal}$
as a function of temperature and supersaturation in air at atmospheric
pressure. Additional measurements for crystals grown in different gases with
various impurity levels as a function of pressure are also needed. As such
measurements accumulate, we hope that the data lead to a better
understanding of ice crystal growth, and of crystal growth dynamics in
general.

\end{document}